\documentclass{birkmult}

\usepackage{mathrsfs}

\title[Spectrum of a Quantum Dot with Impurity in the Lobachevsky Plane]
{On the Spectrum of a Quantum Dot \\ with Impurity in the  Lobachevsky Plane}

\author{P. \v{S}\v{t}ov\'\i\v{c}ek}
\address{Department of Mathematics\\
  Faculty of Nuclear Sciences\\
  Czech Technical University,\\
  Prague, Czech Republic} \email{stovicek@fjfi.cvut.cz}

\author{M. Tu\v{s}ek}
\address{Department of Mathematics\\
  Faculty of Nuclear Sciences\\
  Czech Technical University,\\
  Prague, Czech Republic} \email{tusekmat@fjfi.cvut.cz}

\newtheorem{thm}{Theorem}[section]
\newtheorem{cor}[thm]{Corollary}

\newtheorem{prop}[thm]{Proposition}
\theoremstyle{definition}

\theoremstyle{remark}

\newtheorem*{rem*}{Remark}

\numberwithin{equation}{section}

\newcommand{\ud}{\mathrm{d}}
\newcommand{\parc}[2]{\frac{\partial #1}{\partial #2}}
\newcommand{\abs}[1]{\vert #1 \vert}
\newcommand{\lsz}{\left\lbrace }
\newcommand{\psz}{\right\rbrace }

\newcommand{\ena}[1]{\mathrm{e}^{#1}}
\newcommand{\scp}[2]{\langle#1,#2\rangle}

\newcommand{\N}{\mathbb{N}}
\newcommand{\C}{\mathbb{C}}
\newcommand{\R}{\mathbb{R}}
\newcommand{\Z}{\mathbb{Z}}

\newcommand{\Dom}{\mathop\mathrm{Dom}\nolimits}

\begin{document}

\hyphenation{ana-ly-sis nu-me-ri-ca-lly Mat-he-ma-ti-ca con-si-de-red}

\begin{abstract}
  A model of a quantum dot with impurity in the Lobachevsky plane is
  considered. Relying on explicit formulae for the Green function and
  the Krein $Q-$function which have been derived in a previous work we
  focus on the numerical analysis of the spectrum. The analysis is
  complicated by the fact that the basic formulae are expressed in
  terms of spheroidal functions with general characteristic exponents.
  The effect of the curvature on eigenvalues and eigenfunctions is
  investigated. Moreover, there is given an asymptotic expansion of
  eigenvalues as the curvature radius tends to infinity (the flat case
  limit).
\end{abstract}


\maketitle

\section{Introduction}

The influence of the hyperbolic geometry on the properties of quantum
mechanical systems is a subject of continual theoretical interest for
at least two decades. Numerous models have been studied so far, let us
mention just few of them \cite{com,aco,krr,lisovyy}. Naturally, the
quantum harmonic oscillator is one of the analyzed examples
\cite{bgm,crs}. It should be stressed, however, that the choice of an
appropriate potential on the hyperbolic plane is ambiguous in this
case, and several possibilities have been proposed in the literature.
In \cite{gst}, we have modeled a quantum dot in the Lobachevsky plane
by an unbounded potential which can be interpreted, too, as a harmonic
oscillator potential for this nontrivial geometry. The studied
examples also comprise point interactions \cite{bg} which are
frequently used to model impurities.

A Hamiltonian describing a quantum dot with impurity has been
introduced in \cite{gst}. The main result of this paper is derivation
of explicit formulae for the Green function and the Krein
$Q-$function. The formulae are expressed in terms of spheroidal
functions which are used rather rarely in the framework of
mathematical physics. Further analysis is complicated by the
complexity of spheroidal functions. In particular, the Green function
depends on the characteristic exponent of the spheroidal functions in
question rather than directly on the spectral parameter. In ~fact, it
seems to be possible to obtain a more detailed information on
eigenvalues and eigenfunctions only by means of numerical methods. The
particular case, when the Hamiltonian is restricted to the eigenspace
of the angular momentum with eigenvalue 0, is worked out in \cite{st}.
In the current contribution we aim to extend the numerical analysis to
the general case and to complete it with additional details.

The Hamiltonian describing a quantum dot with impurity in the
Lobachevsky plane, as introduced in \cite{gst}, is a selfadjoint
extension of the following symmetric operator:
\begin{equation*}
  \begin{split}
    & \begin{split}
      H = -\left(\parc{^2}{\varrho^2}
        +\frac{1}{a}\coth\!\left(\frac{\varrho}{a}\right)
        \parc{}{\varrho}+\frac{1}{a^2}
        \sinh^{-2}\!\left(\frac{\varrho}{a}\right)
        \parc{^2}{\phi^2}+\frac{1}{4a^2}\right)
      +\frac{1}{4}\,a^{2}\omega^{2}
      \sinh^2\!\left(\frac{\varrho}{a}\right),
    \end{split} \\
    & \Dom(H) = C^{\infty}_{0}((0,\infty)\times S^1)
    \subset L^{2}\!\left((0,\infty)\times S^1,
      a\,\sinh(\varrho/a)
      \ud\varrho\,\ud\phi\right),
  \end{split}
\end{equation*}
where $(\varrho,\phi)$ are the geodesic polar coordinates on the
Lobachevsky plane and $a$ stands for the so called curvature radius
which is related to the scalar curvature by the formula $R=-2/a^2$.
The deficiency indices of $H$ are known to be $(1,1)$ and we denote
each selfadjoint extension by $H(\chi)$ where the real parameter
$\chi$ appears in the boundary conditions for the domain of
definition: $f(\varrho,\phi)$ belongs to $\Dom(H(\chi))$ if there
exist $f_0,f_1\in\C$ so that $f_1:f_0=\chi:1$ and
\begin{displaymath}
  f(\varrho,\phi) = -\frac{1}{2\pi}\,f_0\log(\varrho)+f_1+o(1)
  \textrm{~}\textrm{~as~}\varrho\to0+
\end{displaymath}
(the case $\chi=\infty$ means that $f_0=0$ and $f_1$ is arbitrary),
see \cite{gst} for details. $H(\infty)$ is nothing but the Friedrichs
extension of $H$. The Hamiltonian $H(\infty)$ is interpreted as
corresponding to the unperturbed case and describing a quantum dot
with no impurity.

After the substitution $\xi=\cosh(\varrho/a)$ and the scaling
$H=a^{-2}\tilde{H}$, we make use of the rotational symmetry (which
amounts to a Fourier transform in the variable $\phi$) to decompose
$\tilde{H}$ into a direct sum as follows
\begin{equation*}
  \begin{split}
    &\tilde{H} = \bigoplus\limits_{m=-\infty}^{\infty}\tilde{H}_{m},\\
    &\begin{split}
      \tilde{H}_{m}
      & = -\parc{}{\xi}\,(\xi^{2}-1)\,\parc{}{\xi}
      +\frac{m^2}{\xi^2-1}+\frac{a^{4}\omega^{2}}{4}\,(\xi^{2}-1)
      -\frac{1}{4}\,,
    \end{split}\\
    & \Dom(\tilde{H}_{m})=C^{\infty}_{0}(1,\infty)
    \subset L^{2}((1,\infty),\ud\xi).
  \end{split}
\end{equation*}

Let us denote by $H_{m}$, $m\in\Z$, the restriction of $H(\infty)$ to
the eigenspace of the angular momentum with eigenvalue $m$. This means
that $H_{m}$ is a self-adjoint extension of $a^{-2}\tilde{H}_{m}$. It
is known (Proposition~2.1 in \cite{gst}) that $\tilde{H}_{m}$ is
essentially selfadjoint for $m\neq0$. Thus, in this case, $H_{m}$ is
the closure of $a^{-2}\tilde{H}_{m}$. Concerning the case $m=0$,
$H_{0}$ is the Friedrichs extension of $a^{-2}\tilde{H}_{0}$. For
quite general reasons, the spectrum of $H_m$, for any $m$, is
semibounded below, discrete and simple \cite{weidmann}. We denote the
eigenvalues of $H_m$ in ascending order by $E_{n,m}(a^{2})$,
$n\in\N_{0}$.

The spectrum of the total Hamiltonian $H(\chi)$, $\chi\neq\infty$,
consists of two parts (in a full analogy with the Euclidean case
\cite{bgl}):
\begin{enumerate}
\item The first part is formed by those eigenvalues of $H(\chi)$ which
  belong, at the same time, to the spectrum of $H(\infty)$. More
  precisely, this part is exactly the union of eigenvalues of $H_{m}$
  for $m$ running over $\Z\setminus\{0\}$. Their multiplicities are
  discussed below in Section~\ref{sec:multiplicities}.
\item The second part is formed by solutions to the equation
  \begin{equation}
    \label{eq:point_levels}
    Q^H(z) = \chi
  \end{equation}
  with respect to the variable $z$ where $Q^{H}$ stands for the Krein
  $Q$-function of $H(\infty)$. Let us denote the solutions in
  ascending order by $\epsilon_{n}(a^{2},\chi)$, $n\in\N_{0}$. These
  eigenvalues are sometimes called the point levels and their
  multiplicities are at least one. In more detail,
  $\epsilon_{n}(a^{2},\chi)$ is a simple eigenvalue of $H(\chi)$ if it
  does not lie in the spectrum of $H(\infty)$, and this happens if and
  only if $\epsilon_{n}(a^{2},\chi)$ does not coincide with any
  eigenvalue $E_{\ell,m}(a^{2})$ for $\ell\in\N_0$ and $m\in\Z$,
  $m\neq0$.
\end{enumerate}

\begin{rem*}
  The lowest point level, $\epsilon_{0}(a^{2},\chi)$, lies below the
  lowest eigenvalue of $H(\infty)$ which is $E_{0,0}(a^2)$, and the
  point levels with higher indices satisfy the inequalities
  $E_{n-1,0}(a^2)<\epsilon_{n}(a^{2},\chi)<E_{n,0}(a^2)$,
  $n=1,2,3,\ldots$.
\end{rem*}

\section{Spectrum of the unperturbed Hamiltonian
  \boldmath{$H(\infty)$}}

Our goal is to find the eigenvalues of the $m$th partial Hamiltonian
$H_{m}$, i.e., to find square integrable solutions of the equation
\begin{equation*}
  H_{m}\psi(\xi)=z\psi(\xi),
\end{equation*}
or, equivalently,
\begin{equation*}
  \tilde{H}_{m}\psi(\xi)=a^{2}z\psi(\xi).
\end{equation*}
This equation coincides with the equation of the spheroidal functions
(\ref{eq:spheroidal_eq}) provided we set $ \mu=\abs{m}$,
$\theta=-a^4\omega^2/16$, and the characteristic exponent $\nu$ is
chosen so that
\begin{equation*}
  \lambda^{m}_{\nu}\!\left(-\frac{a^{4}\omega^{2}}{16}\right)
  = -a^{2}z-\frac{1}{4}\,.
\end{equation*}
The only solution (up to a multiplicative constant) that is square
integrable near infinity is
$S^{\abs{m}(3)}_{\nu}(\xi,-a^{4}\omega^{2}/16)$.

Proposition~\ref{prop:radial_asy} describes the asymptotic expansion
of this function at $\xi=1$ for $m\in\N$. It follows that the
condition on the square integrability is equivalent to the equality
\begin{equation}
  \label{eq:eigen_condition}
  \ena{i(3\nu+1/2)\pi}
  K^{m}_{-\nu-1}\!\left(-\frac{a^{4}\omega^{2}}{16}\right)
  +K^{m}_{\nu}\!\left(-\frac{a^{4}\omega^{2}}{16}\right) = 0.
\end{equation}
Furthermore, in \cite{gst} we have derived that
\begin{equation*}
  S^{0(3)}_{\nu}(\xi,\theta) = \alpha \log(\xi-1)+\beta
  +O((\xi-1)\log(\xi-1))\quad\text{as }\xi\to 1+,
\end{equation*}
where 
\begin{equation*}
  \alpha = \frac{i\tan(\nu\pi)\,\ena{-i(2\nu+1/2)\pi}}
  {2\pi s^{0}_{\nu}(\theta)}
  \left(\ena{i(3\nu+1/2)\pi} K^{0}_{-\nu-1}(\theta)
    +K^{0}_{\nu}(\theta)\right).
\end{equation*}
Taking into account that the Friedrichs extension has continuous
eigenfunctions we conclude that equation (\ref{eq:eigen_condition})
guarantees square integrability in the case $m=0$, too.

\begin{figure}[!ht]\caption{Eigenvalues of the partial Hamiltonian
    $H_{1}$}\label{fig:evodd}
  \begin{center}
  \includegraphics[width=10cm]{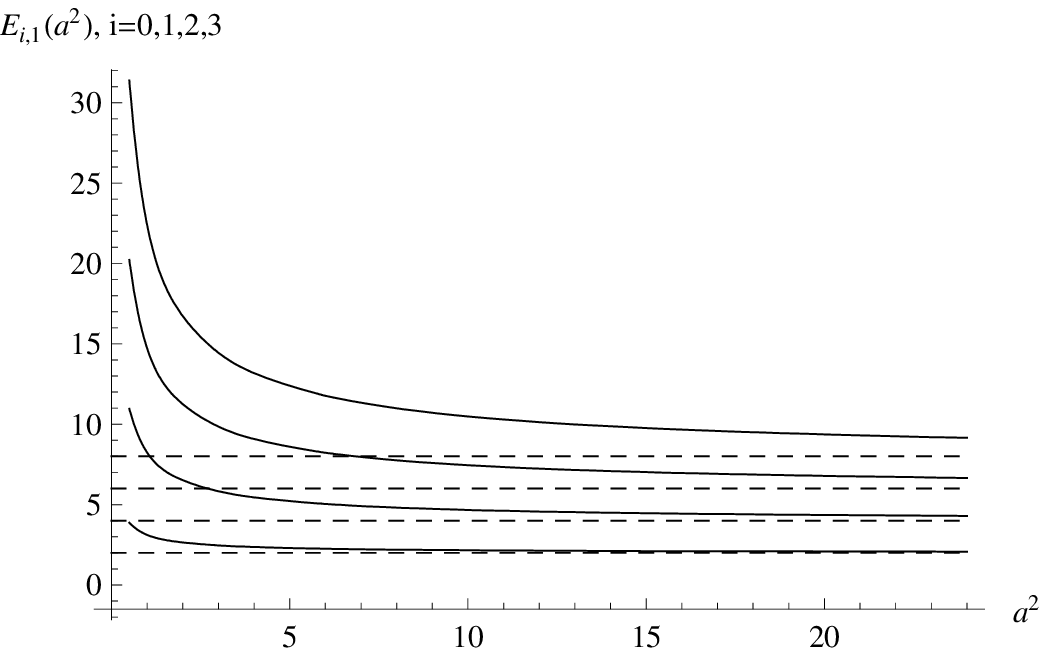}
  \end{center}
\end{figure}

As far as we see it, equation (\ref{eq:eigen_condition}) can be solved
only by means of numerical methods. For this purpose we made use of
the computer algebra system \textit{Mathematica 6.0}. For the
numerical computations we set $\omega=1$. The particular case $m=0$
has been examined in \cite{st}. It turns out that an analogous
procedure can be also applied for nonzero values of the angular
momentum. As an illustration, Figure~\ref{fig:evodd} depicts several
first eigenvalues of the Hamiltonian $H_{1}$ as functions of the
curvature radius $a$. The dashed asymptotic lines correspond to the
flat limit ($a\to\infty$).

Denote the $n$th normalized eigenfunction of the $m$th partial
Hamiltonian $\tilde{H}_{m}$ by $\tilde{\psi}_{n,m}(\xi)$. Obviously,
the eigenfunctions for the values of the angular momentum $m$ and $-m$
are the same and are proportional to
$S^{\abs{m}(3)}_{\nu}(\xi,-a^{4}\omega^{2}/16)$, with $\nu$ satisfying
equation (\ref{eq:eigen_condition}). Let us return to the original
radial variable $\varrho$ and, moreover, regard $\tilde{H}_{m}$ as an
operator acting on $L^{2}(\R^{+},\ud\varrho)$. This amounts to an
obvious isometry
\begin{displaymath}
L^{2}(\R^{+},a^{-1}
\sinh(\varrho/a)\ud\varrho)\to{}L^{2}(\R^{+},\ud\varrho):\
f(\varrho)\mapsto{}a^{-1/2}\sinh^{1/2}(\varrho/a)f(\varrho).
\end{displaymath}
The corresponding normalized eigenfunction of $\tilde{H}_{m}$, with an
eigenvalue $a^{2}z$, equals
\begin{equation}\label{eq:eigenfce}
  \psi_{n,m}(\varrho) = \left(\frac{1}{a}\sinh\left(\frac{\varrho}{a}
    \right)\right)^{\!1/2}
  \tilde{\psi}_{n,m}\!\left(\cosh\left(\frac{\varrho}{a}\right)\right).
\end{equation}
At the same time, relation (\ref{eq:eigenfce}) gives the normalized
eigenfunction of $H_{m}$ (considered on $L^{2}(\R^{+},\ud\varrho)$)
with the eigenvalue $z$. The same Hilbert space may be used also in
the limit Euclidean case ($a=\infty$). The eigenfunctions $\Phi_{n,m}$
in the flat case are well known and satisfy
\begin{equation}\label{eq:flat_ef}
  \Phi_{n,m}\propto\varrho^{\abs{m}+1/2}
  \ena{-\omega \varrho^{2}/4}\,
  {_{1}F_{1}}\!\left(-n,\abs{m}+1,\frac{\omega\varrho^{2}}{2}\right).
\end{equation}

\begin{figure}[b]\caption{The first eigenfunction of the partial
    Hamiltonian $H_{1}$}\label{fig:efce11}
  \begin{center}
  \includegraphics[width=9cm]{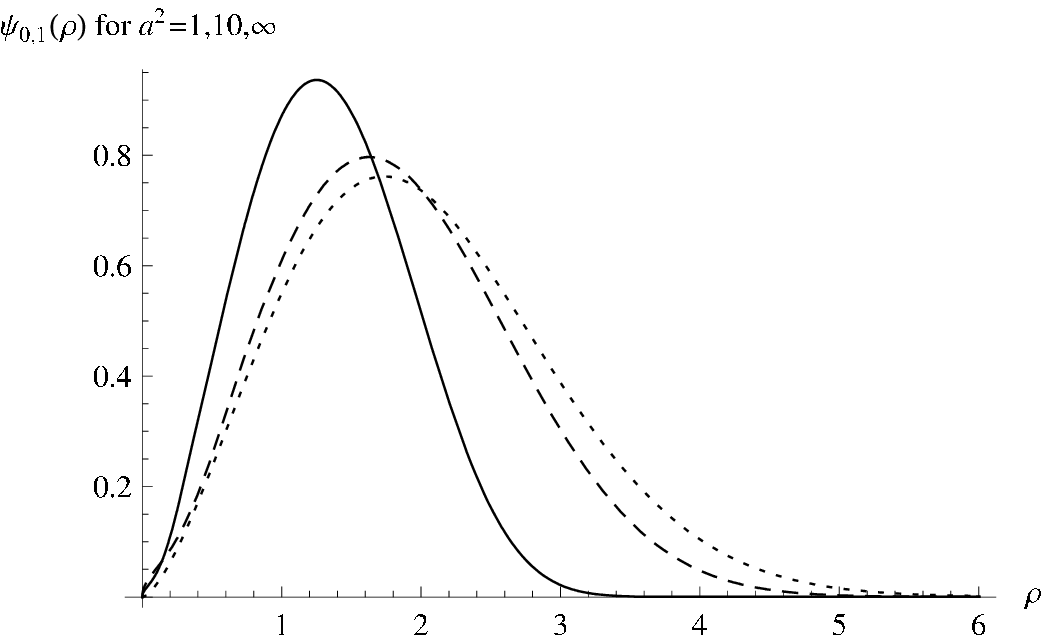}
  \end{center}
\end{figure}

\begin{figure}[t]\caption{The second eigenfunction of the partial
    Hamiltonian $H_{1}$}\label{fig:efce12}
  \begin{center}
  \includegraphics[width=9cm]{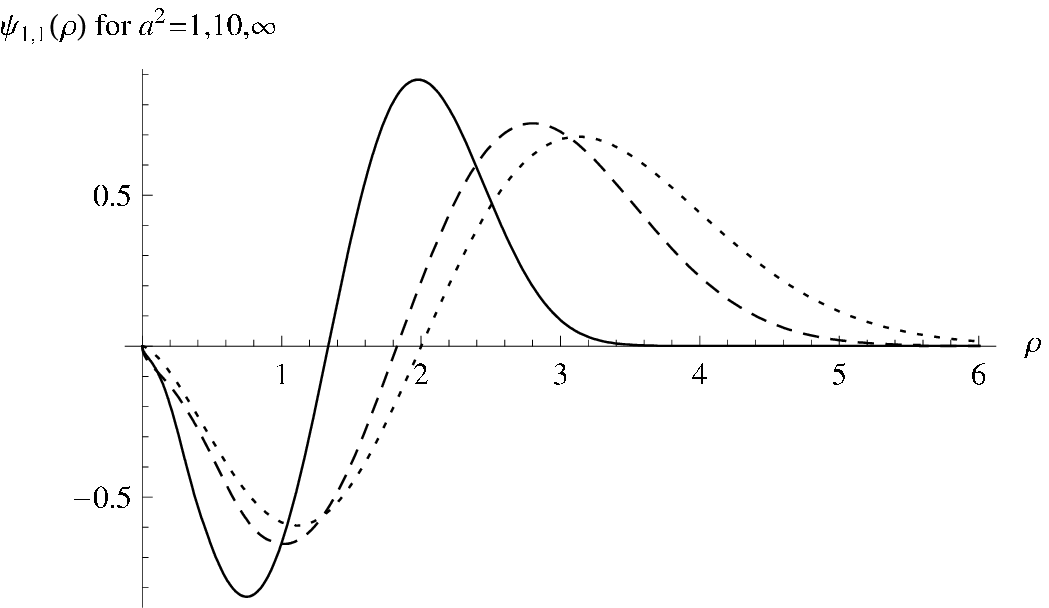}
  \end{center}
\end{figure}

\begin{figure}[t]\caption{The third eigenfunction of the partial
    Hamiltonian $H_{1}$}\label{fig:efce13}
  \begin{center}
  \includegraphics[width=9cm]{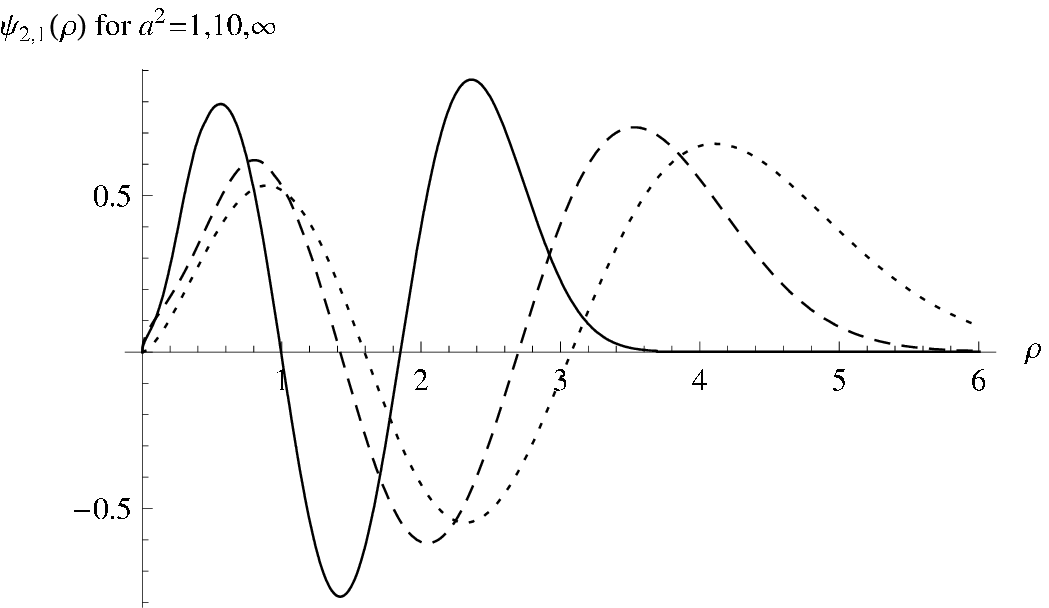}
  \end{center}
\end{figure}

The fact that we stick to the same Hilbert space in all cases
facilitates the comparison of eigenfunctions for various values of the
curvature radius $a$. We present plots of several first eigenfunctions
of $H_{1}$ (Figures~\ref{fig:efce11}, \ref{fig:efce12},
\ref{fig:efce13}) for the values of the curvature radius $a=1$ (the
solid line), $10$ (the dashed line), and $\infty$ (the dotted line).
Again, see \cite{st} for analogous plots in the case of the
Hamiltonian $H_{0}$. Note that, in general, the smaller is the
curvature radius $a$ the more localized is the particle in the region
near the origin.

\section{The point levels}

As has been stated, the point levels are solutions to equation
(\ref{eq:point_levels}) with respect to the spectral parameter $z$.
Since, in general, $Q(\bar{z})=\overline{Q(z)}$ the function $Q(z)$
takes real values on the real axis. Let
$\tilde{H}(\infty)=a^2H(\infty)$ be the Friedrichs extension of
$\tilde{H}$. An explicit formula for the Krein $Q$-function
$Q^{\tilde{H}}(z)$ of $\tilde{H}(\infty)$ has been derived in
\cite{gst}:
\begin{equation*}
  \begin{split}
    Q^{\tilde{H}}(z) = &-\frac{1}{4\pi a^{2}}\left(-\log(2)-2\Psi(1)
      +2\,\Psi s_{\nu}\!\left(-\frac{a^4\omega^2}{16}\right)
      s^{0}_{\nu}\!\left(-\frac{a^4\omega^2}{16}\right)\right)\\
    &+\frac{1}{2 a^{2}\tan(\nu\pi)}\left(\ena{i\pi(3\nu+3/2)}\,
      \frac{K^{0}_{-\nu-1}(-\frac{a^4\omega^2}{16})}{K^{0}_{\nu}
        (-\frac{a^4\omega^2}{16})}-1\right)^{\!-1}
    +\frac{\log{(2a^{2})}}{4\pi a^{2}},
  \end{split}
\end{equation*}
where $\nu$ is chosen so that
\begin{equation*}
  \lambda^{0}_{\nu}\!\left(-\frac{a^4\omega^2}{16}\right)
  = -z-\frac{1}{4}\,.
\end{equation*}
The symbol $K^{0}_{\nu}(\theta)$ stands for the so called spheroidal
joining factor,
\begin{equation*}
  \Psi{}s_{\nu}(\theta)
  := \sum^{\infty}_{r=-\infty}(-1)^{r}a^{0}_{\nu,r}(\theta)\,
  \Psi(\nu+1+2r),
\end{equation*}
where the coefficients $a^{0}_{\nu,r}(\theta)$, $r\in\Z$, come from
the expansion of spheroidal functions in terms of Bessel functions
(for details see \cite[the~Appendix]{gst})), and $s^{0}_{\nu}(\theta)$
is defined by formula (\ref{eq:sum_coef}).  One can obtain the Krein
$Q$-function of $H(\infty)$ simply by scaling
$Q^{H}(z)=a^{2}\,Q^{\tilde{H}}(a^{2}z)$.

\begin{figure}[!h]\caption{The Krein $Q$-function $Q^{H}$ for
    $a^{2}=0.02,~0.2,~1,~5$}\label{fig:Q}
  \begin{center}
  \includegraphics[width=12.5 cm]{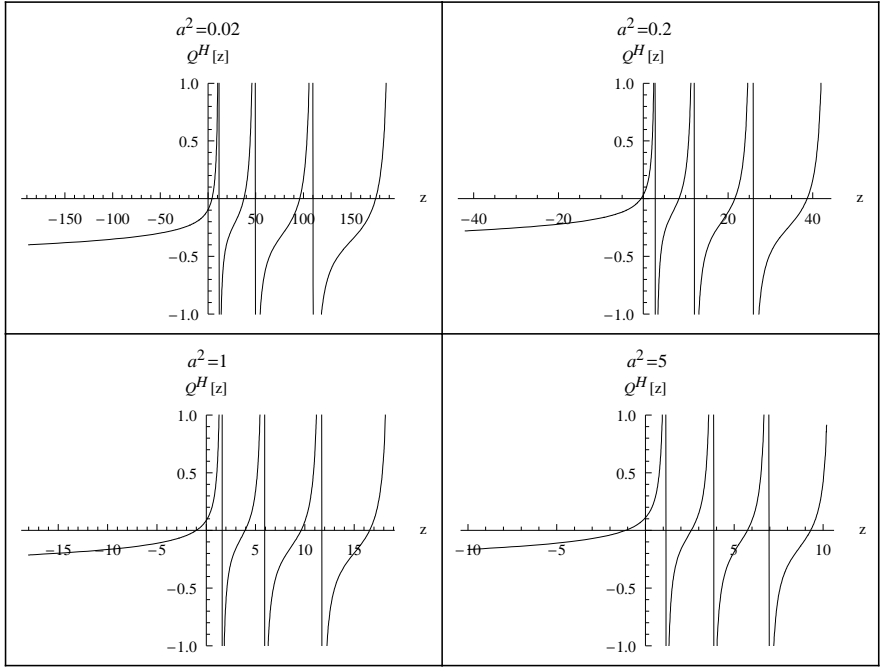}
  \end{center}
\end{figure}

Since we know the explicit expression for the Krein $Q$-function as a
function of the characteristic exponent $\nu$ rather than of the
spectral parameter $z$ itself it is of importance to know for which
values of $\nu$ the spectral parameter $z$ is real.
Propositions~\ref{prop:lambda} and \ref{prop:lambda2} give the answer.
For $\nu\in\R$ and for $\nu$ of the form $\nu=-1/2+it$ where $t$ is
real, the spheroidal eigenvalue
$\lambda^{m}_{\nu}(-a^{4}\omega^{2}/16)$ is real, and so the same is
true for $z$. Moreover, these values of $\nu$ reproduce the whole real
$z$ axis. With this knowledge, one can plot the Krein $Q$-function
$Q^{H}=Q^{H}(z)$ for an arbitrary value of the curvature radius $a$.
Note that for $a=\infty$, the Krein $Q$-function is well known as a
function of the spectral parameter $z$ \cite{gp} and equals (setting
$\omega =1$, $\Psi$ is the logarithmic derivative of the gamma
function)
\begin{equation*}
  Q(z) = \frac{1}{4\pi}\left(-\Psi\!\left(\frac{1-z}{2}\right)
    +\log(2)+2\Psi(1)\right).
\end{equation*}

\begin{figure}[!ht]\caption{Comparison of the Krein $Q$-functions for
    $a^{2}=24$ and $a^{2}=\infty$}\label{fig:Qcompare}
  \begin{center}
  \includegraphics[width=9cm]{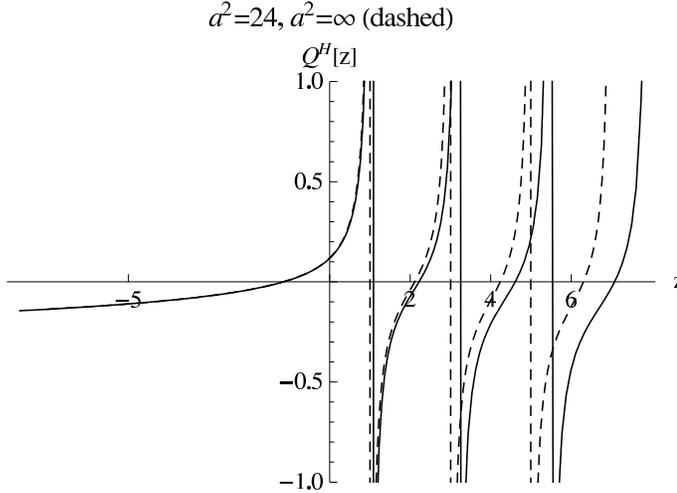}
  \end{center}
\end{figure}

Next, in Figure~\ref{fig:Q}, we present plots of the Krein
$Q$-function for several distinct values of the curvature radius $a$.
Moreover, in Figure~\ref{fig:Qcompare} one can compare the behavior of
the Krein $Q$-function for a comparatively large value of the
curvature radius ($a^{2}=24$) and for the Euclidean case ($a=\infty$).

\begin{figure}[t]\caption{Point levels for H(0)}\label{fig:point_lev1}
 \begin{center}
  \includegraphics[width=10cm]{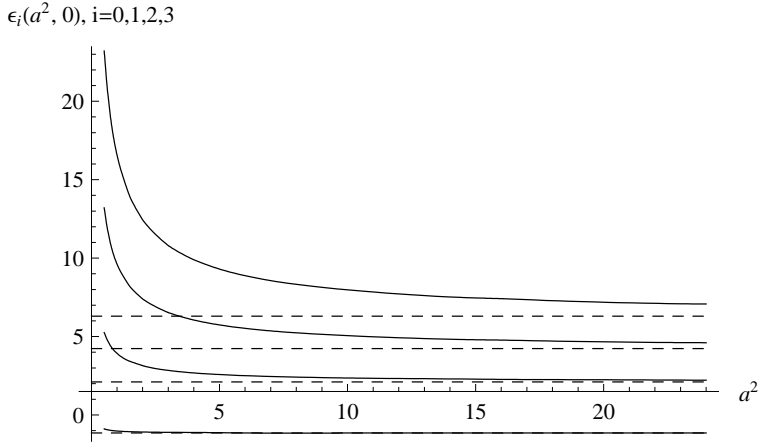}
 \end{center}
\end{figure}

Again, equation (\ref{eq:point_levels}) can be solved only
numerically. Fixing the parameter $\chi$ one may be interested in the
behavior of the point levels as functions of the curvature radius $a$.
See Figure~\ref{fig:point_lev1} for the corresponding plots, with
$\chi=0$, where the dashed asymptotic lines again correspond to the
flat case limit ($a=\infty$). Note that for the curvature radius $a$
large enough, the lowest eigenvalue is negative provided $\chi$ is
chosen smaller than $Q(0)\simeq 0.1195$.

\section{Asymptotic behavior for large values of \boldmath{$a$}}

The $m$th partial Hamiltonian $H_{m}$, if considered on
$L^{2}(\R^{+},\ud\varrho)$, acts like
\begin{equation*} 
  H_{m} = -\parc{^{2}}{\varrho^{2}}
  +\frac{m^{2}-\frac{1}{4}}{a^{2}
    \sinh^{2}(\frac{\varrho}{a})}+\frac{1}{4}\,a^{2}\omega^{2}
  \sinh^{2}\!\left(\frac{\varrho}{a}\right)
  =:-\parc{^{2}}{\varrho^{2}}+V_{m}(a,\varrho).
\end{equation*}
For a fixed $\varrho\neq 0$, one can easily derive that
\begin{equation*}
  V_{m}(a,\varrho) = \frac{m^{2}-\frac{1}{4}}{\varrho^{2}}
  +\frac{1}{4}\omega^{2}
  \varrho^{2}+\frac{\frac{1}{4}-m^{2}}{3a^{2}}
  +\frac{\omega^{2}\varrho^{4}}{12a^{2}}
  +O\!\left( \frac{1}{a^{4}}\right)\quad\textrm{as~}a\to\infty.
\end{equation*}
Recall that the $m$th partial Hamiltonian of the isotropic harmonic
oscillator on the Euclidean plane, $H_{m}^{E}$, if considered on
$L^{2}(\R^{+},\ud\varrho)$, has the form
\begin{equation*}
  H_{m}^{E} := -\parc{^{2}}{\varrho^{2}}+\frac{m^{2}
    -\frac{1}{4}}{\varrho^{2}}
  +\frac{1}{4}\omega^{2}\varrho^{2}.
\end{equation*}
This suggests that it may be useful to view the Hamiltonian $H_{m}$,
for large values of the curvature radius $a$, as a perturbation of
$H_{m}^{E}$,
\begin{equation*}
  H_{m}\sim H_{m}^{E}
  +\frac{1}{12a^{2}}(1-4m^{2}+\omega^{2}\varrho^{4})
  =: H_{m}^{E}+\frac{1}{12a^{2}}U_{m}(\varrho).
\end{equation*}
The eigenvalues of the compared Hamiltonians have the same asymptotic
expansions up to the order $1/a^{2}$ as $a\to\infty$.

Let us denote the $n$th eigenvalue of the Hamiltonian $H^{E}_{m}$ by
$E^{E}_{n,m}$, $n\in\N_{0}$. It is well known that
\begin{equation*}
  E^{E}_{n,m} = (2n+\abs{m}+1)\,\omega
\end{equation*}
and that the multiplicity of $E^{E}_{n,m}$ in the spectrum of $H^{E}$
equals $2n+\abs{m}+1$. The asymptotic behavior of $E_{n,m}(a^2)$ may
be deduced from the standard perturbation theory and is given by the
formula
 \begin{equation}
   \label{eq:ev_asy}
   E_{n,m}(a^{2}) = E_{n,m}^{E}
   +\frac{1}{12a^{2}}\frac{\scp{\Phi_{n,m}}{U_{m}\Phi_{n,m}}}
   {\scp{\Phi_{n,m}}{\Phi_{n,m}}}+O\!\left( \frac{1}{a^{4}}\right)
   \quad\textrm{as~}a\to\infty,
\end{equation}
where $\Phi_{n,m}$ denotes a (not necessarily normalized)
eigenfunction of $H^E_m$ associated with the eigenvalue $E^{E}_{n,m}$
(see (\ref{eq:flat_ef})). The scalar products occurring in formula
(\ref{eq:ev_asy}) can be readily evaluated in $L^{2}(\R^+,\ud\varrho)$
with the help of Proposition~\ref{prop:int}. The resulting formula
takes the form
\begin{equation}
  \label{eq:ev_asy_final}
  E_{n,m}(a^{2}) = (2n+\abs{m}+1)\,\omega
  +\left(2n(n+\abs{m}+1)+\abs{m}+\frac{3}{4}\right)\!\frac{1}{a^{2}}
  +O\!\left(\frac{1}{a^{4}}\right)
\end{equation}
as $a\to\infty$. This asymptotic approximation of eigenvalues has been
tested numerically for large values of the curvature radius $a$.  The
asymptotic eigenvalues for $a^{2}=24$ are compared with the precise
numerical results in Table~\ref{tab:comp}. It is of interest to note
that the asymptotic coefficient in front of the $a^{-2}$ term does not
depend on the frequency $\omega$.

\begin{table}
  \caption{Comparison of numerical and asymptotic results
    for the eigenvalues, $a^{2}=24$}
  \label{tab:comp}
  \begin{center}
    \begin{tabular}{lcccccc}
      & $E_{0,0}$ & $E_{1,0}$ & $E_{2,0}$ & $E_{0,1}$ & $E_{1,1}$ 
      & $E_{2,1}$\\
      \hline
      numerical  & 1.0265 & 3.162 & 5.42  & 2.060 & 4.259 & 6.58  \\
      asymptotic & 1.0268 & 3.169 & 5.46  & 2.058 & 4.258 & 6.59  \\
      error (\%) & -0.03 & -0.22  & -0.74 & 0.10  & 0.02  & -0.15 \\
      \hline
    \end{tabular}
  \end{center} 
\end{table}

\section{The multiplicities}
\label{sec:multiplicities}

Since $H_{-m}=H_m$ the eigenvalues $E_{n,m}(a^{2})$ of the total
Hamiltonian $H(\infty)$ are at least twice degenerated if $m\neq0$.
From the asymptotic expansion (\ref{eq:ev_asy_final}) it follows,
after some straightforward algebra, that no additional degeneracy
occurs and thus theses eigenvalues are exactly twice degenerated at
least for sufficiently large values of $a$.

Applying the methods developed in \cite{bgl} one may complete the
analysis of the spectrum of the total Hamiltonian $H(\chi)$ for
$\chi\neq\infty$. Namely, the spectrum of $H(\chi)$ contains
eigenvalues $E_{n,m}(a^{2})$, $m>0$, with multiplicity 2 if
$Q^{H}(E_{n,m}(a^{2}))\neq\chi$, and with multiplicity 3 if
$Q^{H}(E_{n,m}(a^{2}))=\chi$. The rest of the spectrum of $H(\chi)$ is
formed by those solutions to equation (\ref{eq:point_levels}) which do
not belong to the spectrum of $H(\infty)$. The multiplicity of all
these eigenvalues in the spectrum of $H(\chi)$ equals ~1.

\section*{Appendix: Auxiliary results}

\setcounter{section}{1}
\renewcommand{\thesection}{\Alph{section}}
\setcounter{equation}{0}
\renewcommand{\theequation}{\Alph{section}.\arabic{equation}}

In this appendix we summarize several auxiliary results. Firstly, for
our purposes we need the following observations concerning spheroidal
functions. The spheroidal functions are solutions to the equation
\begin{equation}
  \label{eq:spheroidal_eq}
  (1-\xi^{2})\parc{^2\psi}{\xi^{2}}-2\xi\parc{\psi}{\xi}
  +\left[\lambda^{\mu}_{\nu}(\theta)
    +4\theta(1-\xi^2)-\mu^2(1-\xi^2)^{-1}\right]\psi=0. 
\end{equation}
For the notation and properties of spheroidal functions see \cite{be}.
A detailed information on this subject can be found in \cite{me}, but
be aware of somewhat different notation. A very brief overview of
spheroidal functions is also given in the Appendix of \cite{gst}.

In the last named source, the following proposition has been proved in
the particular case $m=0$. But, as one can verify by a direct
inspection, the proof applies to the general case $m\in\Z$ as well.

\begin{prop}
  \label{prop:lambda}
  Let $\nu,\theta\in\R$, $m\in\Z$. Then
  $\lambda^{m}_{\nu}(\theta)\in\R$.
\end{prop}

\noindent
The following claim is also of interest.

\begin{prop}
  \label{prop:lambda2}
  Let $\nu=-1/2+it$ where $t\in\R$, and $\theta\in\R$, $m\in\Z$. Then
  $\lambda^{m}_{\nu}(\theta)\in\R$.
\end{prop}

\begin{proof}
  Let us recall that the coefficients $a^{m}_{\nu,r}(\theta)$ in the
  series expansion of spheroidal functions in terms of Bessel
  functions satisfy a three term recurrence relation (see
  \cite[the~Appendix]{gst}),
  \begin{equation}
    \label{eq:three_term_a}
    \beta^{m}_{\nu,r}(\theta)a^{m}_{\nu,r-1}(\theta)
    +\alpha^{m}_{\nu,r}(\theta)a^{m}_{\nu,r}(\theta)
    +\gamma^{m}_{\nu,r}(\theta)a^{\mu}_{\nu,r+1}(\theta)
    =\lambda^{m}_{\nu}(\theta)a^{m}_{\nu,r}(\theta).
  \end{equation}
  One may view the set of equations (\ref{eq:three_term_a}), with
  $r\in\Z$, as an eigenvalue equation for $\lambda^{m}_{\nu}(\theta)$
  that is an analytic function of $\theta$. A particular solution is
  fixed by the condition $\lambda^{m}_{\nu}(0)=\nu (\nu+1)$.  Consider
  the set of complex conjugated equations. Since
  $\overline{\beta^{m}_{\nu,r}}=\beta^{m}_{\overline{\nu},r}(\theta)
  =\beta^{m}_{-\nu-1,r}(\theta)$, and the similar is true for
  $\alpha^{m}_{\nu,r}(\theta)$ and $\gamma^{m}_{\nu,r}(\theta)$, it
  holds true that
  \begin{displaymath}
    \beta^{m}_{-\nu-1,r}(\theta)\overline{a^{m}_{\nu,r-1}(\theta)}
    +\alpha^{m}_{-\nu-1,r}\overline{a^{m}_{\nu,r}(\theta)}
    +\gamma^{m}_{-\nu-1,r}(\theta)\overline{a^{\mu}_{\nu,r+1}(\theta)}
    =\overline{\lambda^{m}_{\nu}(\theta)}\,
    \overline{a^{m}_{\nu,r}(\theta)}.
  \end{displaymath}
  Since for each $\nu$ of the considered form,
  \begin{displaymath}
    \lambda^{m}_{-\nu-1}(0)=(-\nu-1)(-\nu)=\nu(\nu+1)
    =\overline{\nu(\nu+1)}=\overline{\lambda^{m}_{\nu}(0)}, 
  \end{displaymath}
  one has
  $\lambda^{m}_{-\nu-1}(\theta)=\overline{\lambda^{m}_{\nu}(\theta)}$.
  Moreover, $\lambda^{m}_{-\nu-1}(\theta)=\lambda^{m}_{\nu}(\theta)$
  in general. We conclude that $\lambda^{m}_{\nu}(\theta)\in\R$.
\end{proof}

Another auxiliary result concerns the asymptotic expansion of the
radial spheroidal function of the third kind.

\begin{prop}
  \label{prop:radial_asy}
  Let $\nu\notin\lsz-1/2+k|\ k\in\Z\psz,\ m\in\N$. Then
  \begin{eqnarray}
    && \label{eq:radial_asy}\hskip-2em
    S^{m(3)}_{\nu}(\xi,\theta) \sim
    \frac{(-1)^{m}2^{m/2-1}\Gamma(m)
      \tan(\nu\pi)}{\pi s^{m}_{\nu}(\theta)\,
      \ena{-i(\nu+3/2)\pi}}\left(K^{m}_{-\nu-1}(\theta)
      +\frac{K^{m}_{\nu}(\theta)}{\ena{i(3\nu+1/2)\pi}}\right)\!
    (\xi-1)^{-m/2}
    \nonumber\\
    && \hskip-2em\textrm{as~}\xi\to 1+.
  \end{eqnarray}
\end{prop}

\begin{proof}
  By the definition of the radial spheroidal function of the third
  kind,
  \begin{equation*}
    S^{m(3)}_{\nu}(\xi,\theta) := \frac{1}{i\cos(\nu\pi)}
    \left(S^{m(1)}_{-\nu-1}(\xi,
      \theta)+i\ena{-i\nu\pi}S^{m(1)}_{\nu}(\xi,\theta)\right),
  \end{equation*}
  and by the relation between the radial and the angular spheroidal
  functions,
  \begin{equation*}
    S^{m(1)}_{\nu}(\xi,\theta)
    = -\frac{\sin(\nu\pi)}{\pi}\,\ena{-i\nu\pi}K^{m}_{\nu}(\theta)\,
    Qs^{m}_{-\nu-1}(\xi,\theta),
  \end{equation*}
  one has
  \begin{equation*}
    S^{m(3)}_{\nu}(\xi,\theta)
    =\frac{i\tan(\nu\pi)}{\pi\ena{-i(\nu+1)\pi}}
    \left(K^{m}_{-\nu-1}(\theta)Qs^{m}_{\nu}(\xi,\theta)
      +\frac{K^{m}_{\nu}(\theta)
        Qs^{m}_{-\nu-1}(\xi,\theta)}{\ena{i(3\nu+1/2)\pi}}\right).
  \end{equation*}
  Using the definition
  \begin{equation*}
    Qs^{m}_{\nu}(\xi,\theta)
    = \sum_{r=-\infty}^{\infty}(-1)^{r}a^{m}_{\nu,r}(\theta)
    Q^{m}_{\nu+2r}(\xi)
  \end{equation*}
  and due to the well known asymptotic expansion for the Legendre
  functions \cite{be},
  \begin{equation*}
    Q^{m}_{\nu}(\xi) \sim (-1)^{m}2^{m/2-1}
    \Gamma(m)(\xi-1)^{-m/2}
    \quad\textrm{as~}\xi\to1+\,,
  \end{equation*}
  one derives that
  \begin{equation*}
    Qs^{m}_{\nu}(\xi,\theta) \sim
    \frac{(-1)^{m}2^{m/2-1}\Gamma(m)}{(\xi-1)^{m/2}s^{m}_{\nu}(\theta)}
    \quad\textrm{as~}\xi\to1+,
  \end{equation*}
  where 
  \begin{equation}\label{eq:sum_coef}
    (s^{m}_{\nu}(\theta))^{-1}
    :=\sum_{r=-\infty}^{\infty}(-1)^{r}a^{m}_{\nu,r}
    (\theta)=\sum_{r=-\infty}^{\infty}(-1)^{r}a^{m}_{-\nu-1,-r}(\theta)
    =(s^{m}_{-\nu-1}(\theta))^{-1}.
  \end{equation}
  Hence $Qs^{m}_{-\nu-1}(\xi,\theta)\sim{}Qs^{m}_{\nu}(\xi,\theta)$ as
  $\xi\to1+$, and one immediately obtains (\ref{eq:radial_asy}).
\end{proof}

Further some auxiliary computations follow that we need for evaluation
of scalar products of eigenfunctions (see (\ref{eq:ev_asy})).

\begin{prop}\label{prop:int}
  Let ${_{1}F_{1}}(a,b,t)$ stand for the Kummer confluent
  hypergeometric function, and $n,m,l\in\N_{0}$. Then
  \begin{equation}\label{eq:hyp_int}
    \begin{split}
      & \int_{0}^{\infty}t^{m+l}
      \ena{-t}{_{1}F_{1}}(-n,1+m,t)^{2}\,\ud t \\
      & =\, (m!)^{2}\sum_{k=\max\lsz 0,n-l\psz}^{n}(-1)^{n+k}
      {n \choose k}\frac{(k+l)!}{(k+m)!}{k+m+l \choose n+m}.
    \end{split}
  \end{equation}
\end{prop}

\begin{proof}
  By definition,
  \begin{displaymath}
      {_{1}F_{1}}(-n,1+m,t)
      := \sum^{n}_{k=0}\frac{(-n)_{k}\,t^{k}}{(1+m)_{k}\,k!}
      = m!\sum_{k=0}^{n}(-1)^{k}{n \choose k}\frac{t^{k}}{(m+k)!}\,.
  \end{displaymath}
  Let us denote the LHS of (\ref{eq:hyp_int}) by $I$. Then the
  integral representation of the gamma function implies
  \begin{equation}
    \label{eq:I}
    I = (m!)^{2}\sum_{j,k=0}^{n}(-1)^{j+k}{n \choose j}
    {n \choose k}\frac{(j+k+m+l)!}{(m+j)!(m+k)!}\,.
  \end{equation}
  Partial summation in (\ref{eq:I}) can be carried out,
  \begin{equation}
    \label{eq:partial_sum}
    \sum_{j=0}^{n}(-1)^{j}{n \choose j}
    \frac{(j+k+m+l)!}{(m+j)!}
    = \frac{\ud^{k+l}}{\ud x^{k+l}}
    \left(x^{k+m+l}(1-x)^{n}\right)\Big|_{x=1}.
  \end{equation}
  Expression (\ref{eq:partial_sum}) vanishes for $k<n-l$ and equals
  \begin{displaymath}
    (-1)^{n}(k+l)!{k+m+l \choose n+m}
  \end{displaymath}
  for $k\geq{}n-l$. The proposition follows immediately.
\end{proof}

\begin{cor}
  In the case $l=0$, (\ref{eq:hyp_int}) takes a particularly simple
  form:
  \begin{equation*}
    \int_{0}^{\infty}t^{m} \ena{-t}{_{1}F_{1}}(-n,1+m,t)^{2}\,\ud t
    = \frac{n!}{(m+n)!}\,.
  \end{equation*}
\end{cor}

\subsection*{Acknowledgments}

The authors wish to acknowledge gratefully partial support of the
Ministry of Education of Czech Republic under the research plan
MSM6840770039 (P.\v{S}.) and from the grant No.~LC06002 (M.T.).

\end{document}